\documentclass[aps,prb,twocolumn,showpacs,groupedaddress]{revtex4}
\usepackage{graphicx}
\begin{document}

\title{Microscopic dynamics and relaxation processes in liquid Hydrogen Fluoride}

\author{
R.~Angelini$^{1}$, P.~Giura$^{1}$, D.~Fioretto$^{2}$
G.~Monaco$^{1}$, G.~Ruocco$^{3}$ and F.~Sette$^{1}$. }
\affiliation{
$^{1}$ European Synchrotron Radiation Facility.
B.P. 220 F-38043 Grenoble, Cedex France.\\
$^{2}$ Universit\'a di Perugia and Istituto Nazionale di Fisica
della Materia, I-06123, Perugia, Italy.\\
$^{3}$ Universit\'a di Roma "La Sapienza" and Istituto Nazionale di Fisica della Materia, I-00185, Roma, Italy.\\
}
\date{\today}
\begin{abstract}

Inelastic x-ray scattering and Brillouin light scattering
measurements of the dynamic structure factor of liquid hydrogen
fluoride have been performed in the temperature range$ T=214\div
283 K$. The data, analysed using a viscoelastic model with a two
timescale memory function, show a positive dispersion of the sound
velocity $c(Q)$ between the low frequency value $c_0(Q)$ and the
high frequency value $c_{\infty \alpha}(Q)$. This finding confirms
the existence of a structural ($\alpha$) relaxation directly
related to the dynamical organization of the hydrogen bonds
network of the system. The activation energy $E_a$ of the process
has been extracted by the analysis of the temperature behavior of
the relaxation time $\tau_\alpha(T)$ that follows an Arrhenius
law. The obtained value for $E_a$, when compared with that
observed in another hydrogen bond liquid as water, suggests that
the main parameter governing the $\alpha$-relaxation process is
the number of the hydrogen bonds per molecule.
\end{abstract}
 \maketitle

\section{Introduction}
To understand how the presence of a relaxation process affects the
dynamics of the density fluctuations in liquids is one of the open
problems in the physics of the condensed matter. Despite the fact
that a large effort has been devoted to shed light on this
subject, the matter is still under debate. In this respect, among
all the relaxations active in a liquid, particular attention has
been paid to relaxation processes of viscous nature which strongly
affect the longitudinal density modes. They include at least two
distinct contributions: a structural (or $\alpha$) and a
microscopic (or $\mu$) process. The $\alpha$-process is associated
to the structural rearrangement of the particles in the liquid and
its characteristic time ($\tau_\alpha$) is strongly temperature
dependent. $\tau_{\alpha}$ can vary several order of magnitude
going from the ps, in the high temperature liquid phase, to $\sim
100~s$ in glass-forming materials at the glass transition
temperature. The $\mu$-process takes its origin from the
oscillatory motion of a particle in the cage of its nearest
neighbors before escaping. Its characteristic time $(\tau_\mu)$ is
shorter than $\tau_\alpha$ and its "strength" is often larger than
the strength of the $\alpha$-process. Other relaxation processes,
beyond the $\alpha$ and the instantaneous processes, associated
with the internal molecular degrees of freedom may be observed in
molecular liquids~\cite{PRL8217761999,PRE650515032002}. The
existence of the $\alpha$ and $\mu$ processes, already introduced
several years ago in a molecular dynamic simulation study on a
Lennard-Jones fluid~\cite{PRA716901973}, has recently been proved
by experiments on liquid
metals~\cite{PRE650312052002,PRE630112102001,PRL8540762000,JPC1280092000}.
In this respect another very important class of liquids to
consider are the hydrogen bonded (HB) liquid systems. In these
compounds indeed, the highly directional hydrogen bond plays a
crucial role in the determination of their microscopic properties.
 Despite the large number of theoretical studies
~\cite{JCP11735582002,JCP10646581997,MP923311997,JCP107101661997,MP93151998,JCP11836392003,
PRL8120801998,JCP11146631999,JCP11290252000}, the way in which the
peculiarities of the hydrogen bond networks affect the static
organization and the dynamical behavior of these compounds is
still a subject of discussion. Many are the parameters related to
the hydrogen bond that must be considered in the description of
the physical properties of these liquids, as for example the
hydrogen bond strength, the spatial network arrangement of the
hydrogen bonds and the number of hydrogen bonds per molecule. From
an experimental point of view, a study of the collective dynamics
as a function of these parameters is extremely important to
clarify the role played by each of them on the physical properties
of HB systems. Among the HB liquids, hydrogen fluoride (HF)
represents one of the most intriguing systems as demonstrated by
the large amount of theoretical study on its static
~\cite{JCP11735582002,JCP10646581997,MP923311997,JCP107101661997,MP93151998,JCP11836392003}
and dynamic ~\cite{PRL8120801998,JCP11146631999,JCP11290252000}
properties. It represents, in fact, a perfect HB model system: it
has a simple diatomic molecule and a very strong hydrogen bond
that determines a linear chain arrangement of the HB network.
Nevertheless, despite its apparent simplicity, only few
experimental data are available because of the very high
reactivity of the material that consequently makes its handling
extremely difficult. In a previous work~\cite{PRL882555032002} we
studied the high frequency dynamics of liquid HF by inelastic
x-ray scattering (IXS) at fixed temperature demonstrating  the
presence of both structural and microscopic relaxation processes.
In the present paper we present an extended study of HF as a
function of the temperature in the liquid phase between $T_B=
292~K$ the boiling point and $T_M= 193~K$ the melting point.
Comparing the results obtained with two different techniques, IXS
and Brillouin light scattering (BLS), we find the presence of a
structural relaxation process in the entire explored temperature
range. The relaxation time $\tau_\alpha(T)$, in the sub-picosecond
time scale, follows an Arrhenius temperature dependence with an
activation energy strictly related to the number of hydrogen
bonds. The paper is organized as follow: Sec.~II is devoted to the
description of the experimental aspects related to the IXS and BLS
measurements of the dynamic structure factor of HF. Sec.~III is
dedicated to the data analysis, in Sec.~IV  the main results are
discussed and finally in Sec.~V  the outcomes of this study are
summarized.

%---------------------------------------------------
\section{Sample environment and experimental set-up}
%---------------------------------------------------

High purity (99.9\%) hydrogen fluoride has been purchased by Air
Products and distilled in the scattering cell without further
purification. The sample cell was made out of a  stainless steel
block, this material is well suited to resist to the chemical
reactivity of HF. To allow the passage of the incident and
scattered beam, two sapphire windows of $250 \mu m$ thickness and
$6~mm$ diameter, have been glued on two holder plates which have
then been screwed to the body of the cell. An o-ring of parofluor
has been applied between the window holders and the cell to
guarantee a good tightness. The whole cell has been
thermoregulated by means of a liquid flux cryostat DC50-K75 Haake.
Further details of the sample cell will be described
elsewhere~\cite{dapubblicare}

\subsection{Brillouin light scattering experiment}\label{BLS}
\begin{figure}
\begin{center}
\includegraphics[width=8.5cm,height=12cm]{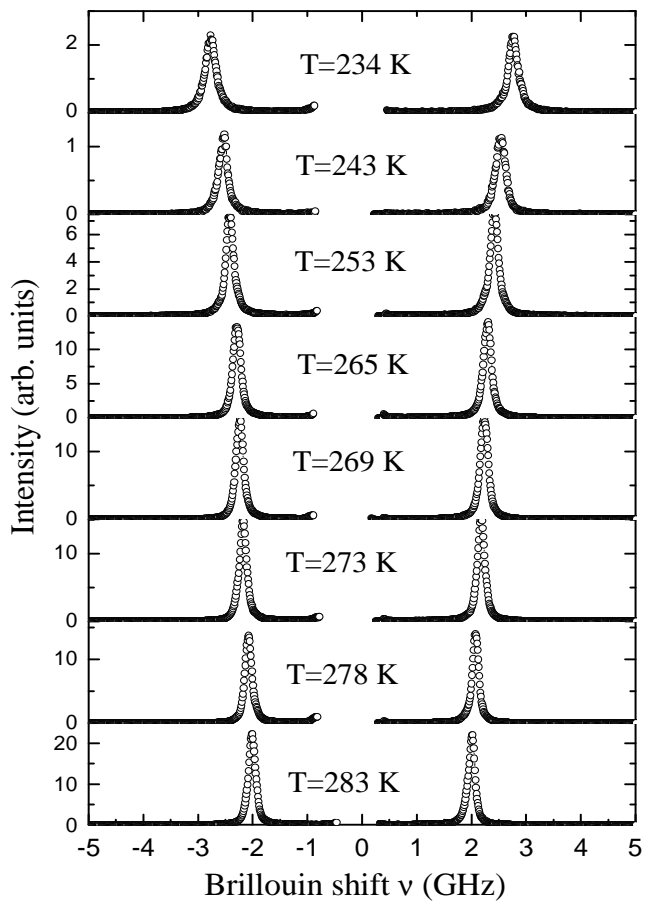}
\caption{Brillouin light scattering spectra of hydrogen fluoride
at the indicated temperatures.} \label{S(Q,w)_light}
\end{center}
\end{figure}

The dynamic structure factor of HF in the GHz range has been
measured by Brillouin light scattering using a Sandercock-type
multi-pass tandem Fabry-Perot interferometer characterized by high
contrast ($ > 5 \cdot 10^{10}$), resolution (FWHM $\approx$ 0.1
GHz) and a finesse of about 100. The wavelength of the incident
radiation was $\lambda = 514.5 nm$ and the light scattered by the
sample was collected in the back-scattering geometry ($\theta =
180^0$). The free spectral range (FSR) was set to 10 GHz, the
integration time was approximately $2.5~s/channel$. The
polarization of the incident light was vertical while the light
scattered by the sample was collected in the unpolarized
configuration. The aim of the present measurement is to determine
the frequency position and width of the Brillouin doublets
associated to the propagation of the sound modes. As the
relaxation time for HF ,in the investigated temperature range, is
in the sub-picosecond region, we do not expect any  evidence of
the mentioned relaxation processes in the GHz range. Thus from the
measured Brillouin peak position and width, it is possible to
extract information about the adiabatic sound velocity $c_0$ and
the kinematic longitudinal  viscosity $\nu_L$.

\subsection{Inelastic x-rays scattering experiment}
The inelastic x-rays experiment has been carried out at the very
high energy resolution IXS beam-line ID16 at the European
Synchrotron Radiation Facility. The instrument consists of a
back-scattering monochromator and five independent analyzers
operating at the (11 11 11) Si Bragg reflection. They are held one
next to the other with a constant angular offset on a 6.5 m long
analyzer arm. The used configuration~\cite{NIMPRB1173391996},
gives an instrumental energy resolution of 1.6 meV full width half
maximum (FWHM) and a Q offset of 3 $nm^{-1}$ between two neighbor
analyzers. The momentum transfer, Q, is selected by rotating the
analyzer arm. The spectra at constant Q and as a function of
energy were measured with a Q resolution of $0.4\ \ nm^{-1}$ FWHM.
The energy scans were performed varying the back-scattering
monochromator temperature with respect to that of the analyzer
crystals. Further details on the beam-line are reported
elsewhere~\cite{NIMPRB1111811996}. Each scan took about 180 min
and each spectrum at fixed Q was obtained by summing up to 3 or 6
scans.

\section{Data reduction}

\begin{figure}
\begin{center}
\includegraphics[width=9cm,height=12cm]{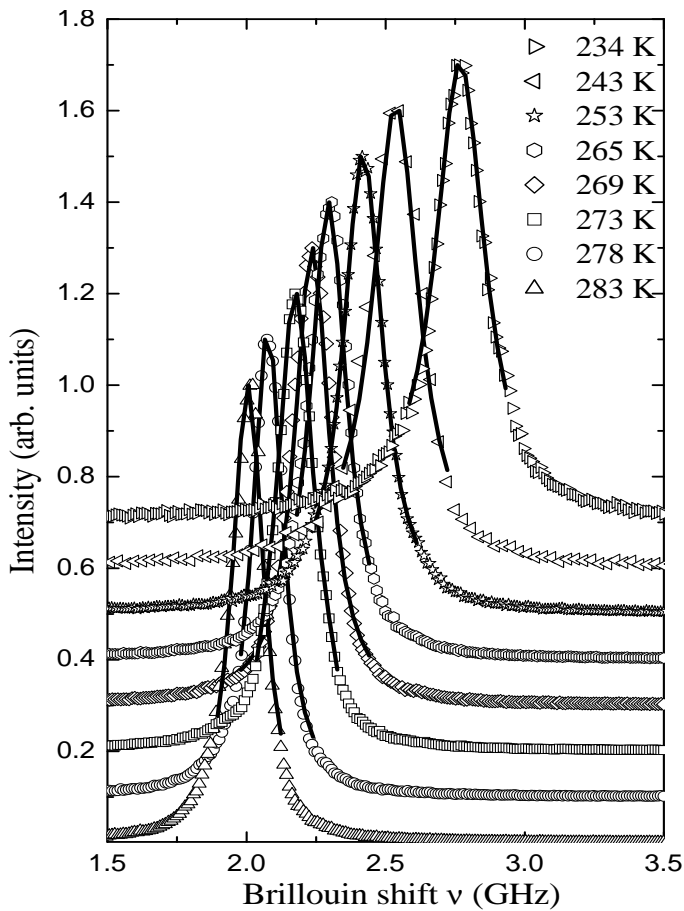}
\caption{Stokes part of the Brillouin light scattering spectra of
HF at the indicated temperatures. The fit (solid line) are
superimposed to the data (open symbols). The curves are shifted on
the y-axis one respect to the other.} \label{stokes(T)}
\end{center}
\end{figure}

\begin{figure}
\begin{center}
\includegraphics[width=8.0cm,height=12cm]{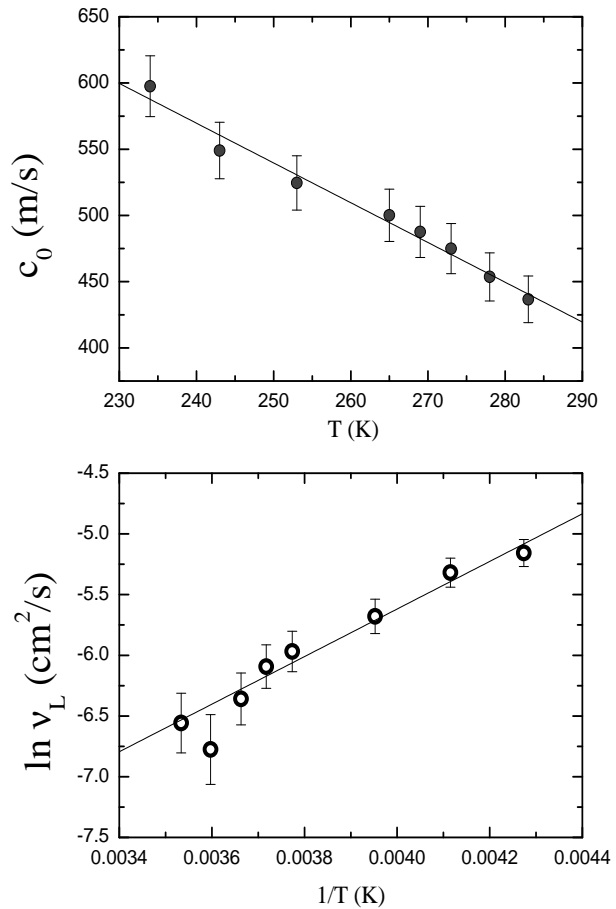}
\caption{(a) Sound velocity $c_0$ and (b) Kinematic longitudinal
viscosity $\nu_L$ from Table~\ref{tab1} as calculated from the
position and the width of the Brillouin light scattering
excitations in back-scattering geometry $(\theta = 180^0)$ (full
circles). The straight lines represent the linear fit to the data.
} \label{c(T)_light}
\end{center}
\end{figure}

\subsection{Brillouin light scattering}
Unpolarized Brillouin spectra collected in a temperature range
between $234~K$ and $283~K$ are shown in Fig.~\ref{S(Q,w)_light}.
The quantities of interest are the position and the width of the
Brillouin peaks directly related to the sound velocity {\bf $c_0$}
and to the kinematic longitudinal viscosity {\bf $\nu_L$} of HF.
In order to extract these two parameters the experimental data
have been fitted in a limited region around the inelastic peaks
with the function obtained by the convolution of the instrumental
resolution $R(\omega)$ with a damped harmonic oscillator (DHO)
function:

\begin{equation}
I(Q,\omega)= R(\omega)\otimes A {{2 \Gamma(Q)
\Omega(Q)^2}\over(\Omega(Q)^2-\omega^2)^2+(2 \omega \Gamma(Q))^2}
\label{fit}
\end{equation}

\noindent where $\Omega = 2 \pi \nu_0$ is the bare oscillation
frequency and $2 \Gamma$ is approximately the full width at half
maximum (FWHM) of the sound excitations. The results of the
fitting procedure are reported in Fig.~\ref{stokes(T)}
superimposed to the experimental data.

\begin{figure*}
\begin{center}
\includegraphics[width=18cm,height=10cm]{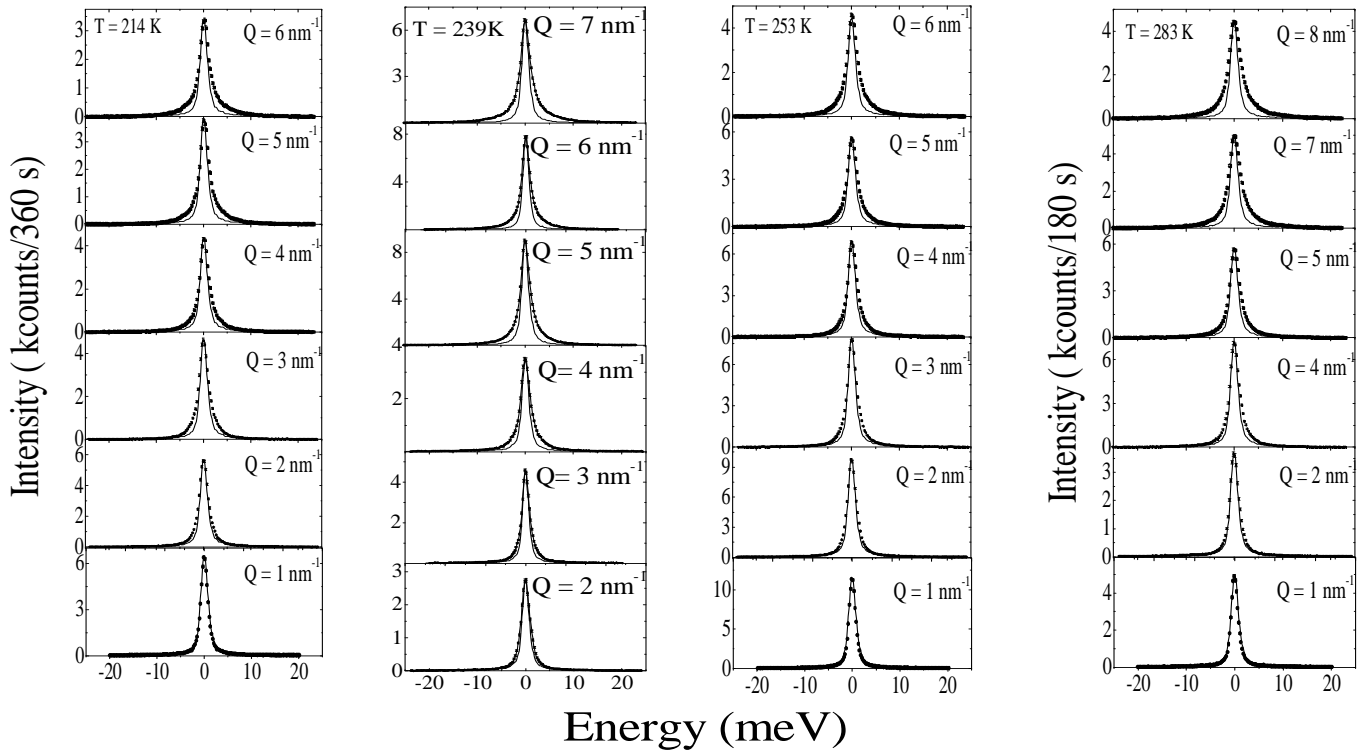}
\caption{ IXS spectra of HF at fixed temperature in the low-Q
region. The raw data (lines + symbol) are plotted together with
the corresponding experimental resolutions (dashed lines).}
\label{S(Q,w)}
\end{center}
\end{figure*}

\begin{table}
\caption{Values of the frequency position $\nu_0$ and of the width
$2 \Gamma$ of the Brillouin doublet as obtained by the fitting
procedure described in the text. The Q values, are also reported
together with the sound velocities $c_0$ and the kinematic
longitudinal viscosities $\nu_L$.}
\begin{ruledtabular}
\begin{center}
\begin{tabular}{ccccccccccccccc}
T & & &$\nu_0$& & & $2 \Gamma /2 \pi$& & & $Q $& & $c_0$& & &$\nu_L$\\
(K)& & &$(GHz)$& & & $(GHz)$& & &$(nm^{-1})$& & $(m/s)$& & &$(cm^2/s)$\\
\hline
234& & & 2.74& & &0.24& & &0.029& &600& & & 0.0058\\
\hline
243& & &2.51& & &0.20& & &0.029& &550& & & 0.0049\\
\hline
253& & &2.39& & &0.14& & &0.029& &530& & & 0.0034\\
\hline
265& & &2.27& & &0.10& & &0.029& &500& & & 0.0026\\
\hline
269& & &2.21& & &0.09& & &0.028& &490& & & 0.0023\\
\hline
273& & &2.15& & &0.07& & &0.028& &480& & & 0.0017\\
\hline
278& & &2.05& & &0.05& & &0.028& &450& & & 0.0011\\
\hline
283& & &1.97& & &0.06& & &0.028& &440& & & 0.0014\\
\end{tabular}
\label{tab1}
\end{center}
\end{ruledtabular}
\end{table}
\noindent By exploiting the relations $\Omega(Q) = c_0 Q$ and
$\nu_L = 2 \Gamma / 2 \pi$, the adiabatic sound velocity $c_0$ and
the kinematic longitudinal viscosity $\nu_L$ are obtained. The
exchanged momentum $Q$ values are determined via the relation
$Q={4 \pi n \over \lambda} sin({\theta \over 2})$, where $ n$ is
the refractive index and $\theta$ is $180^0$ in the used
scattering geometry. The temperature dependent refractive index,
$n(T)$, has been obtained by using the Clausius-Mossotti relation:
\begin{equation}
{n(T)^2 -1 \over n(T)^2+2} = {4 \over 3} \pi \rho_n(T) \alpha
\label{refractive}
\end{equation}

\noindent where $\rho_n$ is the number density and $\alpha$ the
optical polarizability  of the HF molecule. The latter quantity,
$\alpha$, has been obtained using the values of $n(T \approx 293
K)$~\cite{handbook} and $\rho(T \approx 293 K)$~\cite{landolt} and
assuming no  temperature dependence for this parameter  which
turns out to be $\alpha = (9.606 \pm 0.008)~\AA^3$. The data for
$n(T)$ have then been obtained at each temperature by using
$\alpha$ and $\rho(T)$ whose expression is given by
~\cite{landolt}

\begin{equation}
\rho (T) = \rho_0 + A\cdot T  \label{density}
\end{equation}
\noindent with $\rho_0=(1.616 \pm 0.003)~g/cm^3$ and $A=-(2.25 \pm
0.01) \cdot 10^{-3}~g/cm^3K$. The derived values of the sound
velocity $c_0$ and the kinematic longitudinal viscosity $\nu_L$
are reported in Tab. I and shown in Fig.~\ref{c(T)_light}. The
quantity $c_0(T)$ follows a linear behavior characterized by a
temperature dependence well represented by the equation:

\begin{equation}
c_0(T) = c_0 + B \cdot T \label{velocity}
\end{equation}

\noindent with $c_0=(1290\pm 40)~m/s$ and $B=(- 3.0\pm 0.1)~m/sK$.
The same procedure has been applied to derive the kinematic
longitudinal viscosity $\nu_L$ for which the linear fit provides a
temperature behavior described by the relation:

\begin{equation}
ln \nu_L(T) = C + {D \over T}
\end{equation}

\noindent with $C= (- 13.5 \pm 0.7)~cm^2/s$ and $D= (1960\pm
170)~cm^2K/s$. All the values of Q, of the fit parameters and of
the calculated $\nu_L$ and $c_0$, are reported in
Table~\ref{tab1}.

%--------------------------------------
\subsection{Inelastic x-rays scattering}
%--------------------------------------

The IXS measurements, performed to probe the dynamics of HF in the
mesoscopic regime, are compared to the BLS results of the previous
section in order to characterize the transition from the
hydrodynamic regime to the mesoscopic. In this case the
$S(Q,\omega)$ has been studied at four temperatures in the range
214-283 K  at T = 214 K, T = 239 K, T = 254 K and T = 283 K as a
function of the wave vector Q. It has been varied between $1 \div
15 nm^{-1}$ for all the studied temperatures excepted for T = 239
K, where it has been selected between $2 \div 31 nm^{-1}$. Each
energy scan took 180 min and each spectrum at fixed Q was obtained
by summing up to 6 or 3 scans. We report in Fig.~\ref{S(Q,w)} an
example of the measured spectra at the investigated temperatures
and at the indicated momentum transfer (dotted line); they are
compared with the instrumental resolution aligned and scaled to
the central peak (full line).

%-----------------------------
\subsubsection{Markovian approach}
%-----------------------------
\begin{figure}
\begin{center}
\includegraphics[width=8cm,height=11cm]{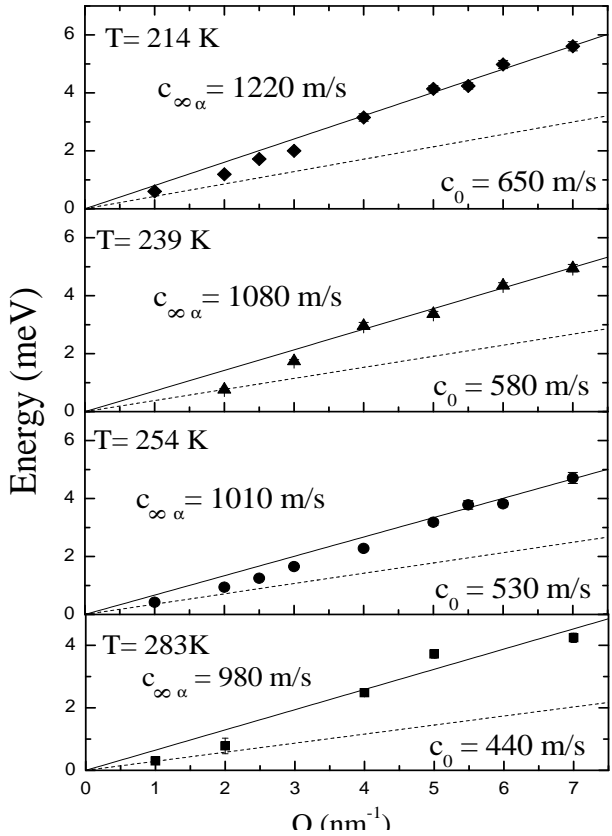}
\caption{Dispersion curves at the indicated temperatures. The
upper full lines are the linear fits to the high-Q data. The lower
dashed lines indicate the adiabatic sound velocity as measured by
Brillouin light scattering as shown in Sec.~\ref{BLS}.}
\label{w(Q)_T}
\end{center}
\end{figure}

\begin{figure}
\begin{center}
\includegraphics[width=8cm,height=11.3cm]{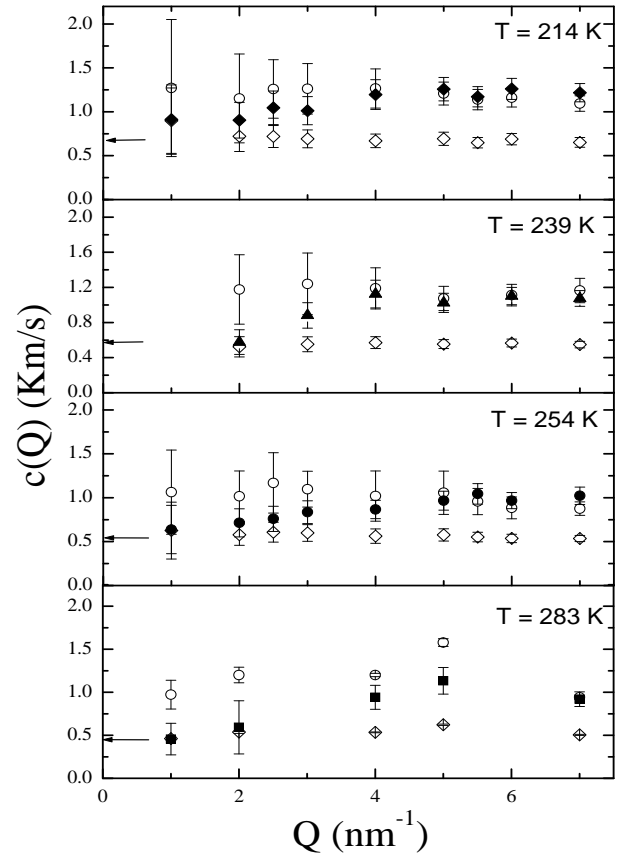}
\caption{Q-dependence of the sound velocities $c_0(Q)$ (open
diamonds) and $c_{ \infty \alpha}(Q)$ (open circles) from a
viscoelastic analysis, together with $c(Q)$ (full symbols) from
Fig.~\ref{w(Q)_T}. The value of the adiabatic sound velocity,
$c_0$, is indicated by the arrow. }\label{c(Q)_T}
\end{center}
\end{figure}
A first raw analysis of the spectra has been done in terms of the
Markovian approximation in the memory function
approach~\cite{balucani}. In this approach the $S(Q,\omega)$ is
expressed as:

\begin{equation}
S(Q,\omega)=I(Q){{\omega_0(Q)}^2 M^{\prime}(Q,\omega)\over
[\omega^2-\omega_0(Q)^2-\omega M^{\prime
\prime}(Q,\omega)]^2+[\omega
M^{\prime}(Q,\omega)]^2}\label{viscoelastic}
\end{equation}

\noindent where $\omega_0(Q)^2= (K_BT / m S(Q))Q^2$ is the
normalized second frequency moment of $S(Q,\omega)$, $K_B$ is the
Boltzmann constant, $m$ is the mass of the molecule and
$M^{\prime}(Q,\omega)$, $M^{\prime \prime}(Q,\omega)$ are
respectively the real and the imaginary part of the Laplace
transform of the memory function $M(Q,t)$. In the Markovian
approximation, the decay of the memory function is faster than any
system time scale and is modelled with a $\delta$ function in the
time domain~\cite{balucani} in such a way that
Eq.~\ref{viscoelastic} reduces to a DHO. To  fit our data we used
this function plus a Lorentzian to take into account the finite
width of the quasi elastic central peak . The detailed balance and
the convolution with the instrumental resolution have also been
taken into account during the fitting procedure. One of the
parameters we are interested in, is $\Omega(Q)$ which corresponds
to the frequency of the sound modes. Its dispersion curve (i.e.
its Q-dependence)is shown in Fig.~\ref{w(Q)_T} at low $Q$ for the
four analysed temperatures. The data show a common behaviour in
the entire investigated T range, namely a linear dependence in the
$Q$ range $4-5 \div 7~nm^{-1}$, with a slope corresponding to
sound velocities higher than the adiabatic values $c_o$ measured
by BLS and reported in the previous section. In addition for lower
Q, the apparent sound velocity $c(Q)$ determined by IXS seems to
show a transition from the low frequency value $c_o$ to the higher
value. This result provides the necessary information to extend
what recently observed in liquid HF at $T =
239~K$~\cite{PRL882555032002} to a wider temperature range. The
increase of $c(Q)$ with increasing Q, in fact, is interpreted as
due to the presence of a relaxation process, the structural
relaxation, already observed in HF at $T=239~K$ and still present
in the whole explored temperature region.

\subsubsection{Viscoelastic approach}
\begin{figure}
\begin{center}
\includegraphics[width=8.5cm,height=11cm]{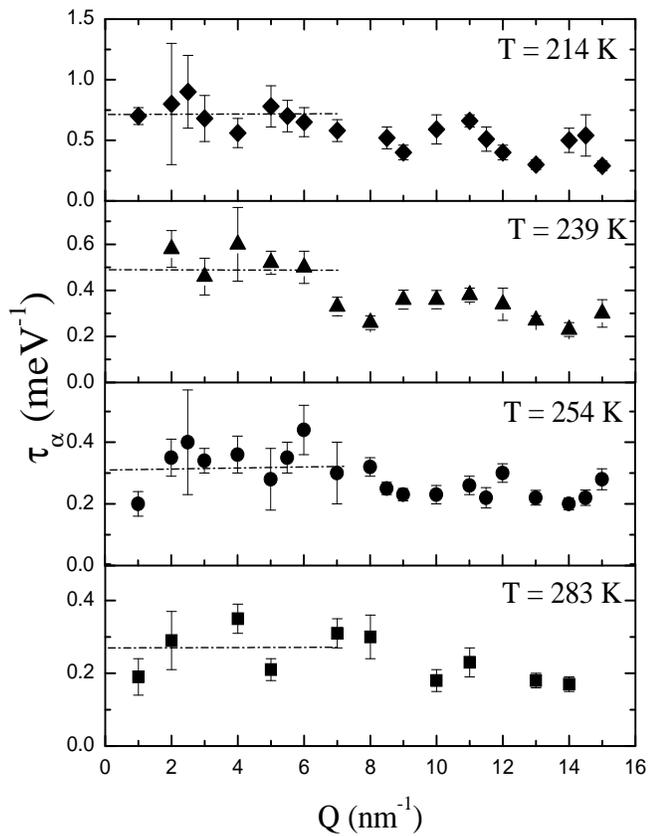}
\caption{ Q-dependence of the relaxation times $\tau_{\alpha}(Q)$
from the viscoelastic analysis at the indicated temperatures. A
constant fit in the low Q region (dashed line) is also reported.
}\label{tau(Q)_T}
\end{center}
\end{figure}

\begin{figure}
\begin{center}
\includegraphics[width=8cm,height=11cm]{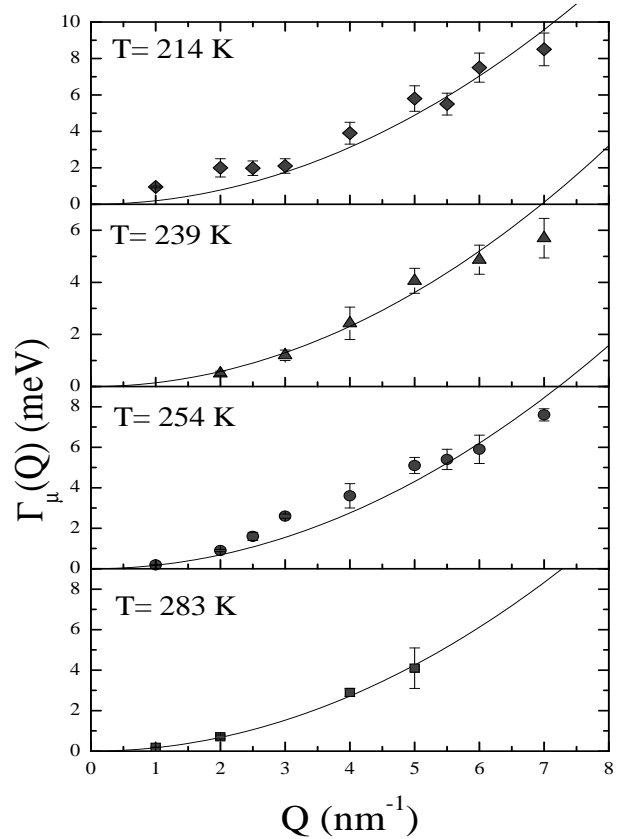}
\caption{Q-dependence of the parameter $\Gamma_{\mu}(Q)$ (full
circles) in liquid HF at the analyzed temperatures from a
viscoelastic analysis. The full lines are the parabolic fits to
the low-Q data. } \label{gamma(Q)_T}
\end{center}
\end{figure}

The existence of a relaxation process with a characteristic time
$\tau$ in the range of the probed sound waves (i.e. such that
$\Omega \tau \simeq 1$), as evident from the dispersion of
$\Omega(Q)$, calls for a more refined choice of the memory
function with respect to the Markovian approximation. To describe
the effect of this relaxation in $S(Q,\omega)$, we use the memory
function based on the viscoelastic model. In this approach we
describe a two relaxation process scenario with a memory function
$M(Q,t)$ given by the sum of an exponential decay contribution and
a $\delta$-function~\cite{PRL882555032002} :

\begin{equation}
M(Q,t)= \Delta_\alpha^2(Q)e^{-{t/ \tau_\alpha(Q)}} +
\Gamma_\mu(Q)\delta(t) \label{memory2}
\end{equation}

\noindent where
$\Delta_\alpha^2(Q)=[c_{\infty\alpha}(Q)^2-c_0(Q)^2]Q^2$, is the
strength of the $\alpha$ process and $\Gamma_\mu(Q)= \Delta_\mu^2
\tau_\mu(Q)$. As in pure HF at $T= 239~K$~\cite{PRL882555032002}
in fact, one expects a structural process described by an
exponential decay and a microscopic process, very fast respect to
the investigate time scale~\cite{PRE6055051999} described by a
$\delta-$function. This approach has been successfully applied in
the past  to describe the dynamics of simple
liquids~\cite{JCP11422592001} and liquid
metals~\cite{PRL892555062002,PRE650312052002,PRE630112102001,PRL8540762000,JPC1280092000}.
The thermal contribution in the memory function has been neglected
being the value of the specific heats ratio $\gamma$ close to 1.
The experimental data have been fitted to the convolution of the
experimental resolution function with the dynamic structure factor
model given by the combination of Eq.~\ref{viscoelastic}
and~\ref{memory2}.

\noindent
The Q dependence of the fit parameters
$c_{\infty\alpha}(Q)$, $c_0(Q)$, $\tau_\alpha(Q)$ and
$\Gamma_\mu(Q)$ is described in the following .

In Fig.~\ref{c(Q)_T} the Q behavior of the sound velocities is
shown in the low Q region and for all the investigated
temperatures. Both he infinite $c_{\infty \alpha}(Q)$ and zero
$c_0(Q)$ frequency limiting values, as obtained from the
viscoelastic fit, are reported (open symbols). The comparison with
the apparent sound velocity $c(Q) = \Omega(Q)/Q$ (full symbols) as
derived from the dispersion curve of Fig.~\ref{w(Q)_T}, shows the
transition of c(Q) from $c_0(Q)$ to $c_{\infty \alpha}(Q)$. The
consistency between the two independent analysis strongly suggests
that this transition is governed by the $\alpha$-relaxation
process in the entire temperature range.

The Q dependence of the relaxation time $\tau_{\alpha}(Q)$ is
reported in Fig.~\ref{tau(Q)_T} at the four investigated
temperatures and in the $Q$ range $1 \div 15~nm^{-1}$. It shows a
constant behavior, within the error bars, in the low Q region and
a very week decrease at increasing Q as already observed in
water~\cite{PRE6055051999} and many other systems~\cite{balucani}.
The fit of the data in the $1\div 7~nm^{-1}$ range yields values
 $\tau_{\alpha}(Q \rightarrow 0)$ reported in Tab.~\ref{tab2}. In
Fig.~\ref{gamma(Q)_T} the Q dependence of the strength of the
microscopic relaxation, $\Gamma_{\mu}(Q)$, is reported at the four
analyzed temperatures. The data show a quadratic behavior and have
been fitted with a parabolic function

$$\Gamma_{\mu}(Q)=D Q^2$$

\noindent The values of the parameter $D$ are reported in
Tab~\ref{tab2} as a function of the temperature.

\begin{figure}
\begin{center}
\includegraphics[width=9cm,height=7cm]{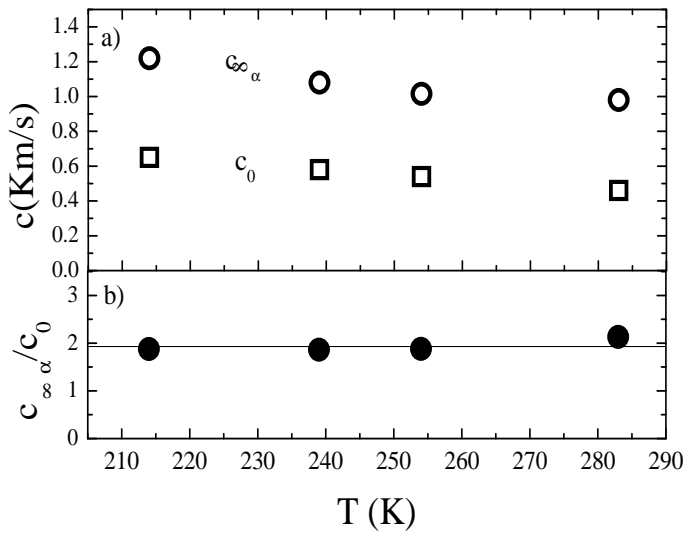}
\caption{(a) Behavior of the sound velocities in $HF$ as a
function of the temperature : $c_0$ (open squares), $c_{\infty
\alpha}$ (open circles). (b) Sound velocities ratio
$c_{\infty\alpha}/c_0$ (symbols); the straight line represents a
constant fit to the data.} \label{cinfsuczero}
\end{center}
\end{figure}

\begin{figure}
\begin{center}
\includegraphics[width=9cm,height=7cm]{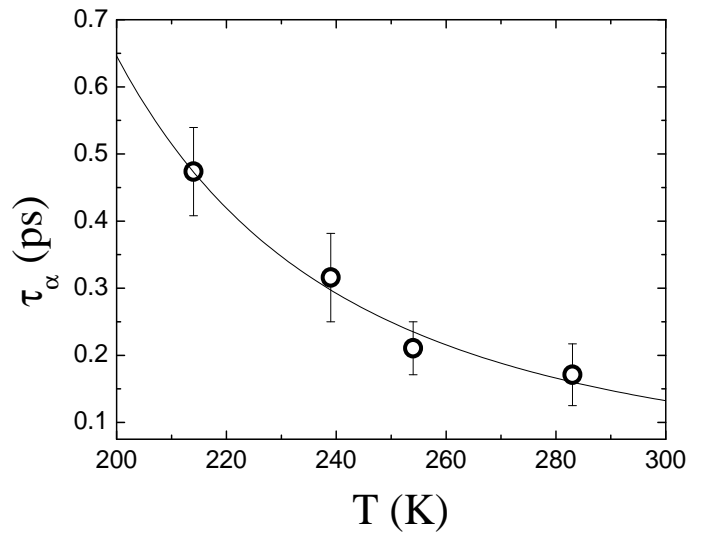}
\caption{T dependence of the low-Q extrapolation of the relaxation
times $\tau_{\alpha}(Q)$ of Figure~\ref{tau(Q)_T} together with
the Arrhenius fit (full line) of Eq.~\ref{Arrhenius}.
}\label{tau(T)}
\end{center}
\end{figure}
\begin{table}

\caption{Low Q values of the parameters used to calculate the
kinematic longitudinal viscosity $\nu_L(0)$ of
Eq.~\ref{viscosity}: $c_0(0)$ from Eq.~\ref{velocity};
$c_{\infty\alpha}(0)$ from the linear fit of Fig.~\ref{w(Q)_T};
$\tau(0)$ and $\Gamma$ derived from the viscoelastic analysis as
described in the text.}
\begin{ruledtabular}
\begin{center}
\begin{tabular}{cccccccccccccccc}
T & & & $c_0(0)$ & & & $c_{\infty \alpha}(0)$ & & & $\tau_\alpha(0)$ & & & $D$ & & & $\nu_L(0)$\\
$(K)$& & &$(m/s)$ & & & $(m/s)$& & &$(ps)$& & &$(cm^2/s)$&&& $(cm^2/s)$\\
\hline
214& & & 650 & & & 1220 & & & 0.47 & & & 0.0030 & & & 0.007\\
\hline
239& & & 580 & & & 1080 & & & 0.32 & & & 0.0022 & & & 0.004\\
\hline
254& & & 530 & & & 1010 & & & 0.21 & & & 0.0026 & & & 0.003\\
\hline
283& & & 480 & & & 980  & & & 0.17 & & & 0.0026 & & & 0.003\\
\end{tabular}
\end{center}
\end{ruledtabular}
\label{tab2}
\end{table}

%--------------------
\section{Discussions}
%--------------------

\begin{figure}
\begin{center}
\includegraphics[width=8.6cm,height=7cm]{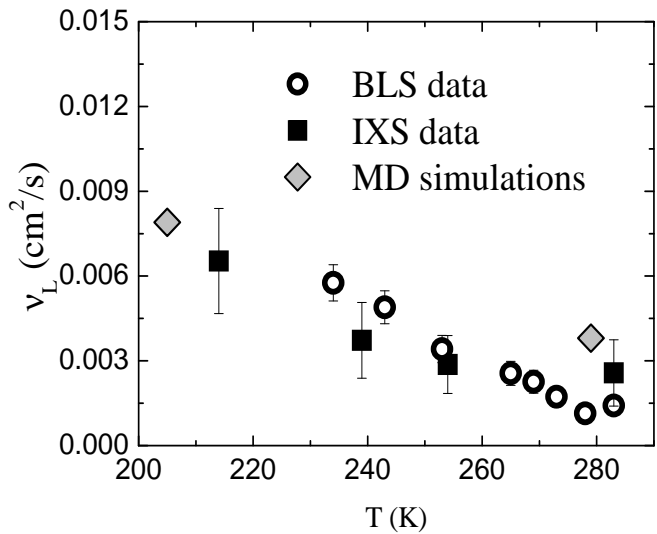}
\caption{ Temperature dependence of the kinematic longitudinal
viscosity: from IXS as calculated through Eq.~\ref{viscosity}
(full squares), from the Brillouin light scattering values of
Table~\ref{tab1} (open circles) and from molecular dynamic
simulations at $T=205 K$ and $T = 279 K$ (full
diamonds)~\cite{JCP11290252000}. }\label{viscosity_xrays}
\end{center}
\end{figure}

This section is dedicated to the discussion of the temperature
dependence of the low Q behavior of the different parameters
analysed in the previous paragraphs. These parameters fully
characterize the collective dynamics of our system in the
mesoscopic regime. The values of $c_0$, $c_{\infty\alpha}$ and
$c_{\infty\alpha}/c_o$ at the four investigated temperatures are
reported in Fig.~\ref{cinfsuczero}. As shown in
Fig.~\ref{cinfsuczero}(b) the ratio $c_{\infty\alpha}/c_o$ is
temperature independent in all the explored T-range and it is
close to two as in the case of water~\cite{PRL77831996}. The
temperature dependence of the structural relaxation time in the $Q
\rightarrow 0$ limit has been deduced from Fig.~\ref{tau(Q)_T}.
Here the low $Q$ part of $\tau_{\alpha}(Q)$ has been fitted using
a constant function. The obtained values are reported in
Fig.~\ref{tau(T)} on a linear scale as a function of the
temperature. In the explored temperature range, the
$\tau_{\alpha}(T)$ behavior is well described by the Arrhenius law
(full line):

\begin{equation}
\tau_{\alpha}(T) = \tau_0 e^{{E_a \over K_B T}}\label{Arrhenius}
\end{equation}

\noindent with an activation energy $E_a = (1.9 \pm 0.2)~kcal/mol$
and $\tau_0 = (6 \pm 2 )10^{-15} s$. The temperature dependence in
the limit $Q \rightarrow 0$ of the last fit parameter $D$ is
reported in Tab.~\ref{tab2}. It yields values which appear to be
temperature independent being $D = (0.170\pm 0.025)~meV/nm^{-2} =
(2.6 \pm 0.4)\cdot 10^{-3}~cm^2/s $. This result is consistent
with previous findings according to which the microscopic
relaxation is a temperature independent
process~\cite{PRE6055051999}. Using the low Q values
$\tau_{\alpha}(0)$ of Tab.~\ref{tab2} together with the low Q
extrapolations of the other parameters (see Tab.~\ref{tab2}), it
is possible to calculate the kinematic longitudinal viscosity
$\nu_L$ from the relation~\cite{balucani}:

\begin{equation}
\nu_L = \tau_{\alpha}({0})( c_{\infty \alpha}^2(0) - c_0^2(0)) +
{D\over 2Q^2}\label{viscosity}
\end{equation}

\noindent where $c_0$ is the adiabatic sound velocity measured by
Brillouin light scattering as discussed in previous section. The
derived values for $\nu_L$ are reported in
Fig.~\ref{viscosity_xrays}: they are found to be consistent with
the hydrodynamic data reported in Tab.~\ref{tab1} . This
equivalence gives further support to the validity of the employed
viscoelastic model. In the same figure we also report two
viscosity data determined by molecular dynamics (MD) simulation
simulations. A recent MD study of the transport coefficients
(longitudinal and shear viscosity, thermal diffusivity and
conductivity) of hydrogen fluoride~\cite{JCP11290252000} provides
two values for the longitudinal viscosity $\eta_L$, one at $T =
205 K$, $\eta_L(T=205K) = 0.91 \cdot 10^{-2} g/cm s $ and the
other at $T = 279 K$, $\eta_L(T=279K) = 0.38 \cdot 10^{-2} g/cm s
$. They are reported in Fig.~\ref{viscosity_xrays} after rescaling
for the density of Eq.~\ref{density} according to the relation
$\nu_L(T)=\eta_L(T)/\rho(T)$, they are quite consistent with the
experimental data.

\begin{figure}
\begin{center}
\includegraphics[width=8cm,height=7cm]{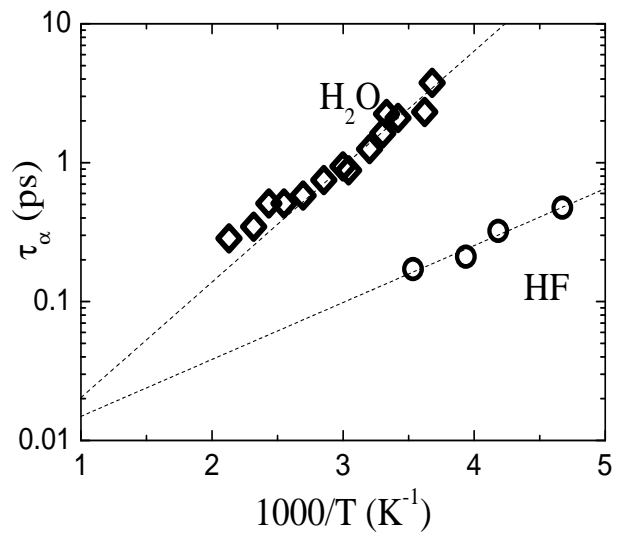}
\caption{ Arrhenius plot for the relaxation time as obtained from
viscoelastic analysis of the dynamical structure factor for water
(diamonds)~\cite{PRE6055051999} and hydrogen fluoride (circles);
The dashed lines indicate the best linear fit to the data and
their slop give an activation energy of $3.8~kcal/mol$ for water
and $1.9~kcal/mol$ for HF.} \label{tauArrhenius}
\end{center}
\end{figure}
\noindent In Fig.~\ref{tauArrhenius} we report, on an Arrhenius
plot, the comparison between the relaxation times for hydrogen
fluoride and for water~\cite{PRE6055051999}. The activation energy
found in water, constant in the examined temperature range, was
$E_a = (3.8 \pm 0.6 )~kcal/mol$ while the one for hydrogen
fluoride is $E_a = (1.9 \pm 0.2)~kcal/mol$ as previously
discussed. It is worthwhile to relate the values of the activation
energies to the different networks present in the two liquids.
While hydrogen fluoride forms linear chains with one hydrogen bond
on average for each molecule, the preferred arrangement of water
is the three-dimensional tetraedric structure with two hydrogen
bonds for each molecule. If we indicate with $n_{HB-H_2O}$ and
$n_{HB-HF}$ the number of hydrogen bonds for $H_2O$ and $HF$
respectively and $E_{a-H_2O}$, $E_{a-HF}$ the activation energies
for the two liquids, we see that they satisfy the ratio:

\begin{equation}
{E_{a-H_2O} \over n_{HB-H_2O} } \approx {E_{a-HF} \over
n_{HB-HF}}\label{newequation}
\end{equation}

\noindent In previous studies on water~\cite{JCP6050251974} the
activation energy has been associated with that of the H-bond
$(\approx5 kcal/mol)$~\cite{pauling}. The result of
Eq.~\ref{newequation} strengthens the idea that the structural
relaxation process involves the H-bond networks of the system and
it seems also to suggest that in this case the activation energy
of the process is related to the number of H-bonds to make and
break and not only to the strength of each H-bond.

\section{Conclusions}

We have presented inelastic Brillouin light and inelastic x-rays
scattering measurements of liquid hydrogen fluoride, a prototype
of the class of hydrogen bonded liquid systems, in a temperature
range comprised between $214~K$ and $283~K$. We demonstrated that
the collective dynamics of liquid HF is characterized by a
structural relaxation process in the sub-picosecond time scale. In
the explored temperature region this relaxation process affects
the collective dynamics in a Q range between $1 \div 7 nm^{-1}$.
An accurate analysis in terms of the viscoelastic model in the
memory function approach allowed to extract and determine the
temperature dependence of the parameters describing the  dynamics
at microscopic level. The relaxation time related to the
structural relaxation process follows an Arrhenius temperature
behavior with an activation energy $E_a$ which, compared with the
value previously measured in liquid water, enables to establish a
connection between $E_a$ and the number of hydrogen bonds per
molecule of the specific system.

\begin{acknowledgments}
We acknowledge R.~Verbeni for assistance during the measurements
C.~Henriquet for the design, development and assembly of the
hydrogen-fluoride cell, C.~Lapras for technical help and C.
Alba-Simionesco, M.C. Bellissent-Funel and J.F. Legrand for useful
discussions.
\end{acknowledgments}

\bibliographystyle{apsrev}

\end{document}